\documentclass[
 reprint,
superscriptaddress,
groupedaddress,
unsortedaddress,
runinaddress,
frontmatterverbose, 
preprint,
showpacs,preprintnumbers,
nofootinbib,
nobibnotes,
bibnotes,
 amsmath,amssymb,
 10pt,
 aps,
]{revtex4-1}
\usepackage{dcolumn}
\usepackage{bm}
\usepackage{hyperref}
\usepackage[mathlines]{lineno}
\usepackage[latin1]{inputenc}
\usepackage{physics}
\usepackage{graphicx}
\usepackage{color}
\usepackage{bm}
\usepackage{dsfont}
\usepackage{longtable}
\usepackage{amsmath} 
\usepackage{amsfonts}
\usepackage{amssymb}
\usepackage{hyperref}
\usepackage{makeidx}
\usepackage{float}
\usepackage[toc,page]{appendix}
\usepackage{natbib}

\begin{document}
\title{Game-theoretic relevance of temporal quantum correlations}
\author{Debsuvra Mukhopadhyay}
\email{debsip@bose.res.in}
\affiliation{Satyendra Nath Bose National Centre for Basic Sciences, Block-JD, Sector-III, Salt Lake, Kolkata-700 098, India}

\begin{abstract}
In this work, we explore a new direction by complementing the game-theoretic applications of nonlocal correlations through appropriately formulated games using temporal quantum correlations. In the context of Bayesian games, we show the way temporal correlations can be utilized, thereby leading to quantum strategies that are impossible to simulate in a classical world. Furthermore, by educing some intriguing characteristics of projective measurements performed between two temporally separated measurements on a given system, we also construct some specific types of Biased Bayesian games. Finally, we impart a new dimension to the cooperative Bayesian Nonlocal Game introduced in [\href{http://www.nature.com/ncomms/2013/130703/ncomms3057/full/ncomms3057.html}{Nat. Comm. 4, 2057 (2013)}] through expedient inclusion of temporal correlations in the EPR-Bohm setting. 
\end{abstract}


\maketitle

\section{Introduction}
In recent years, several categories of quantum games (\cite{1}-\cite{23},\cite{27}) have been constructed, all of which indubitably reveal the superior potentiality of certain strategies accessible only in the quantum domain. Harnessing nonlocal resources for game-theoretic implementations (\cite{7}-\cite{23}) has been a primary cynosure of research in this direction. For instance, a number of interesting nonlocal games (\cite{7}-\cite{11},\cite{14}-\cite{18},\cite{23}) have been studied by setting up close associations with Bell's inequalities. Nonlocality has been one of the most captivating hallmarks of quantum mechanics that has been employed extensively for information processing tasks, as for example, in teleportation (\cite{28}-\cite{31}) and cryptographic protocols (\cite{32},\cite{33}). Tersely stated, nonlocality concerns the phenomenon whereby two spatially separated agents sharing a pair of entangled qubits can generate correlations that cannot be reproduced by any local realistic theory. Hence the correlations are said to be nonlocal. This feature was first explicitly demonstrated by John Bell in the EPR-Bohm setting through the formulation of testable inequalities (\cite{34},\cite{36},\cite{38}) characterizing a specific type of constraint which local correlations are bound to satisfy but which is violated by some quantum correlations. \\

A particular category of nonlocal games known as Bayesian CHSH games was originally discussed by Brunner and Linden (\cite{18}), and very recently, by Anna Pappa et al (\cite{22}), albeit in a different setting, through which the connection between Bayesian games (\cite{25},\cite{26}) and Bell Nonlocality was made much more categorical. In our work, we transcend the relevance of spatial correlations and highlight the role of temporal correlations with a similar rationale of probing its game-theoretic applications in the context of Bayesian games. Just like a violation of the Bell-CHSH inequality (\cite{35},\cite{37}) paves the way for characterizing nonlocal correlations between two spatially separated measurements, violations of the Leggett-Garg inequality (LGI) (\cite{39}-\cite{49},\cite{51}-\cite{53},\cite{58},\cite{59},\cite{61},\cite{62}) and the temporal CHSH inequality (\cite{50},\cite{60},\cite{63}) decidedly exhibit the existence of certain types of temporal correlations in the quantum domain arising due to the inherent invasiveness of quantum mechanical measurements. Central to the game-theoretic application of temporal quantum correlations is the appreciation of the crucial distinction between ideal non-invasive measurements realizable in the classical domain and inherently invasive measurements characteristic of the quantum world. Nonlocal cooperative Bayesian games, as discussed in \cite{18}, exploit violations of the CHSH inequality (\cite{35}) to demonstrate the existence of \textit{super-classical} payoffs. By \textit{super-classical} payoffs, we imply expected payoffs of the players that lie beyond the range of payoffs feasible in the classical domain. Such payoffs can be secured by operating with nonlocal correlations (characterizable through the CHSH inequality) which, in turn, necessitates a cooperation between two active participants.\\

In a similar vein, we formulate a cooperative \textit{temporal Bayesian game} by deploying violations of the Leggett-Garg Inequality to illustrate the existence of \textit{super-classical} payoffs. It is a derivative of the XOR-type game (\cite{11}) so devised as to entail a bijective connection between the expected payoff and the three-term expression pertaining to the third-order Leggett-Garg Inequality (\cite{39}). The inequality requires measuring a dichotomic physical quantity $Q$ (having a discrete spectrum of values $\pm1$) at three different times $t_1$, $t_2$, and $t_3$ respectively ($t_1<t_2<t_3$) and computing the complete set of two-time correlation functions $C_{ij}=\expval{Q(t_i)Q(t_j)}$ to forge an expression of the form $\Delta_L=C_{12}+C_{23}-C_{13}$. The LGI mandates that for a specific class of measurements that do not cause any disturbance to the fundamental ontic state (\cite{54}-\cite{58}) of the system $\Delta_L$ must be bounded above by unity:
\begin{align}
C_{12}+C_{23}-C_{13}\leqslant 1 \label{e1}
\end{align}
Although initially proposed with the intention of testing our classical intuition of the macroscopic world, the LGI is interesting enough for microscopic objects, mostly due to an intimate connection between violations of the same and the behaviour of a physical system under measurement (\cite{61}). We formally introduce the connection between Bayesian Games and the Leggett-Garg Inequality in section \ref{s3}, but only after we recapitulate, in a nutshell, the essence of the two-party Bayesian framework in section \ref{s2}, as this model would serve as the quintessential bedrock in setting up the temporal version.\\

Subsequently, with the objective of making temporal Bayesian games a little bit more interesting, we expand the sphere of participants in the customary Bayesian setting and provide a suitable connotation to the objectives of those actively participating in it. This modification involves insetting a conflict of interest between two asymmetric groups of players. It is easy to appreciate, as has already been stated earlier, that the primary incentive behind constructing quantum Bayesian games (\cite{18},\cite{22}) is to underscore how quantum correlations can help to outmanoeuvre classical strategies. However, one might also be tempted into considering games in which the ``effective" quantumness of the correlations may be viewed as the ultimate deciding factor in settling a conflict of interests. By ``effective" quantumness, we make reference to the perceived nature of the correlations as ascertained through the machinery of suitable correlation inequalities (like the LGI). Consequently, in sections \ref{s4} and \ref{s5}, we construct a few ``biased" games of this sort, in the context of which we demonstrate some interesting attributes of ``intervening" projective measurements that make it possible, at least superficially, to quell the quantumness of temporal correlations, thereby precluding the possibility of achieving super-classical payoffs, even when the games are played quantum mechanically and the correlations are manifestly quantum. Thus, the purpose of this novel formulation is not to simply throw light on the quantum-classical divide as regards the quantitative nature of correlations. Instead, we are motivated by the urge to probe quantum correlations in its own right without obligating a comparison, qualitative or quantitative, with classical correlations. Taking a cue from the observations pertaining to a third-party intervention in the temporal Bayesian game, we examine, in section \ref{s6}, the effect of intervening measurements in a spatio-temporal setting and delineate its game-theoretic applicability. This feature is unravelled in the familiar (nonlocal) Bayesian CHSH game (\cite{18}) fittingly reworked by introducing a time-gap between the measurements effected by Alice and Bob on their respective qubits, in order that the relevance of a third-party intervention at an intermediate instant may be investigated. 
\section{General Formulation of Bayesian Games} \label{s2}

To get started, we lay out, succinctly, the general setting in which quantum Bayesian games (\cite{18},\cite{22}) have hitherto been formulated. The game consists of two players Alice and Bob along with an outside party usually called the referee. The latter is, however, not an active participant in the game. The modus operandi of the game is fairly simple, with the referee designated the task of asking questions to each of Alice and Bob who are supposed to provide answers in response. The respective sets of questions from which the referee chooses his questions for Alice and Bob are denoted $\mathcal{K}$ and $\mathcal{L}$, while the sets of possible answers that Alice and Bob can come up with are denoted $\mathcal{R}$ and $\mathcal{S}$ respectively. Contingent on the pair of answers $\{r,s\}$ provided to a given pair of questions $\{k,l\}$, the referee allocates a payoff to the two players. Let us denote the corresponding payoff function as $\mu(r,s|k,l)$, where $r\in\mathcal{R}, s\in\mathcal{S}, k\in\mathcal{K}, l\in\mathcal{L}$. That is, mathematically,
\begin{align}
\mu : (\mathcal{K},\mathcal{L}) \cross (\mathcal{R},\mathcal{S}) \rightarrow \mathbb{R}
\end{align} As is always the case for cooperative games, higher the payoff, better for the two players engaged in a cooperation. The question-answer rounds are repeated for an ideally infinite number of times with the questions following a well-defined long-time probability distribution represented by $\xi(k,l)$ $(\geqslant0)$ and the corresponding responses the conditional distribution $\alpha(r,s|k,l)$ $(\geqslant0)$. The probabilities are obviously assumed to be normalized, i.e. $\sum_{k\in\mathcal{K},l\in\mathcal{L}}\xi(k,l)=1$ and $\sum_{r\in\mathcal{R},s\in\mathcal{S}}\alpha(r,s|k,l)=1$ $\forall k\in\mathcal{K},l\in\mathcal{L}$. The average payoff $\Pi(\xi, \alpha)$ can then be easily computed:

\begin{align}
\Pi(\xi,\alpha)&=\sum_{k\in\mathcal{K},l\in\mathcal{L}}\xi(k,l)\sum_{r\in\mathcal{R},s\in\mathcal{S}}\alpha(r,s|k,l)\mu(r,s|k,l)\\
&=\sum_{k\in\mathcal{K},l\in\mathcal{L}}\xi(k,l)\Tr(A_{kl}^{T}M_{kl}) \label{e3}
\end{align}
where $A_{kl}$ and $M_{kl}$ are $n(\mathcal{R})\cross n(\mathcal{S})$ matrices with elements given by $[A_{kl}]_{rs}=\alpha(r,s|k,l)$ and $[M_{kl}]_{rs}=\mu(r,s|k,l)$. $[M_{kl}]$ is referred to as the payoff matrix. Here, by $n(\mathcal{A})$ we denote the cardinality of the set $\mathcal{A}$. The arguments $\xi$ and $\alpha$ of the expected payoff function show its dependence on the probability distributions of the choices of both the questions as well as the responses.\\ 
\begin{figure}[ht!]
\centering
\includegraphics[width=8cm, height=6cm]{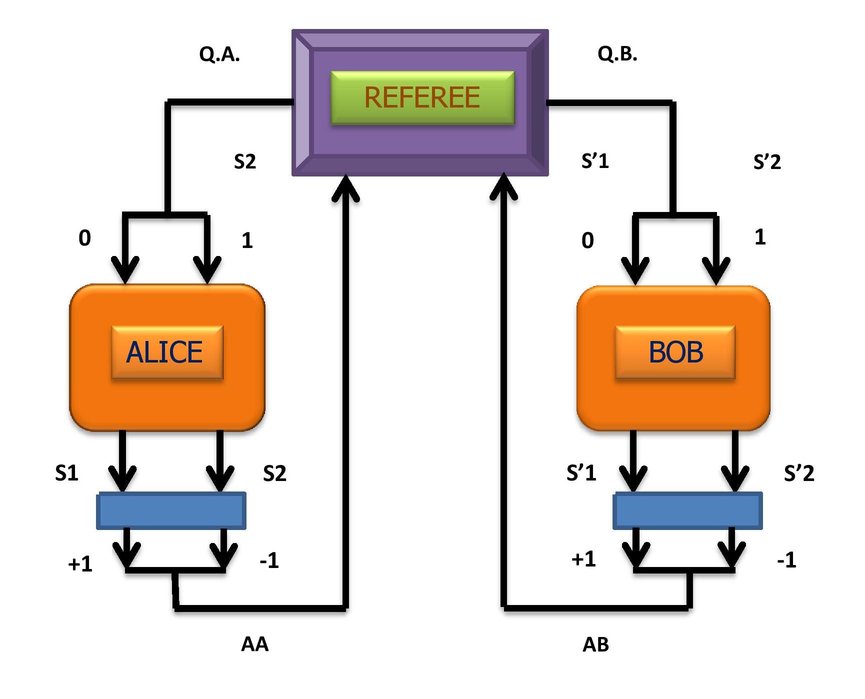}
\caption{\small The customary Bayesian setting for the XOR-type game is depicted above. $S1, S2$ and $S'1,S'2$ represent the possible measurement settings of Alice and Bob respectively. What setting would be chosen by either party is determined by the question posed to the latter. $Q.A.\in\{0,1\}$ and $Q.B.\in\{0,1\}$ are the respective questions posed by the referee to Alice and Bob. The outcomes of measurements performed by the two parties are two-valued, $\pm1$. $AA$ and $AB$ are the answers provided by Alice and Bob depending on the measurement outcomes respectively observed.}
\end{figure}

Such games are called games with incomplete information because, while the players have prior knowledge of the distributions $\alpha$, $\xi$ and the payoff function $\mu$, they cannot communicate between themselves in order to know each other's questions and answers. However, they can connive together before the start of the game and decide upon a strategy that could help them in their quest for a high payoff. Such a strategy is usually referred to as shared randomness in the classical context, while in the quantum setting a predetermined strategy involves a suitable choice of shared quantum state and measurement schemes. \\

If the players deploy independent strategies based purely on the questions asked to them, then $\alpha(r,s|k,l)$ is of the product form: $\alpha_{M}(r|k)\alpha_{M}'(s|l)$, where $\alpha_{M}(r|k)$ and $\alpha_{M}'(s|l)$ are the marginals of the joint distribution. However, it is also possible to adopt correlated strategies based on some classical advice (the source of shared randomness), parametrized by $\lambda$ and delivered to both the players by an external advisor. The advice $\lambda$ can be distributed according to some distribution $\rho(\lambda)$. Each player can then choose a strategy contingent on the question posed to him and on the common advice. So, for classical correlated strategies, we have
\begin{align}
\alpha(r,s|k,l)=\sum_{\lambda}\rho(\lambda)\alpha_M(r|k,\lambda)\alpha_M'(s|l,\lambda) \label{el}
\end{align}

 In the quantum context, it is possible to generate correlations that do not admit this decomposition. For instance, the EPR-Bohm nonlocal correlations are inconsistent with such a restrictive decomposition rule. This inconsistency is, however, not endemic to spatial correlations only - there exist temporal quantum correlations as well that do not adhere to this restriction. The incompatibility is manifested through violations of suitable correlation inequalities characterizing some algebraic constraints on correlations of the form \ref{el}. The relevance of any correlation inequality (such as the CHSH inequality) that places a constraint on a certain category of correlations can be made explicit in the Bayesian framework by appropriately mapping the questions to measurement settings and the answers to the observed outcomes (\cite{18}). As an immediate consequence of the correlation inequality, the expected payoff arising from the corresponding category of correlations is forced to satisfy an inequality (equivalent to the original correlation inequality) of the following form:
\begin{align}
\Pi(\xi,\alpha)\leqslant \Pi_{max}
\end{align}

In the following section, we establish a connection between a variant of the XOR-type game discussed by Brunner and Linden (\cite{18}), and the Leggett-Garg Inequality (\cite{39}) which encapsulates a restriction on temporal correlations resulting from ideal noninvasive measurements characteristic of classical systems.

\section{The Bayesian LGI Game} \label{s3}

To construct this game, rather than considering just two individual players we envisage two groups of players calling them $A$ and $B$ respectively, with each group comprising a total of $N$ members. We label these players $(A_1,A_2,A_3,...,A_N)$ and $(B_1,B_2,B_3,...,B_N)$. Also, we now have a group of $N$ referees, $R$, with the $i^{th}$ referee $R_i(1\leqslant i\leqslant N)$ assigned the task of interrogating the $i^{th}$ pair of players $\{A_i, B_i\}$ - $A_i$ is questioned earlier and $B_i$ is questioned at a later time. The purpose behind tweaking the usual scenario of the nonlocal CHSH game will soon become clear.\\

\begin{figure}[ht!]
\centering
\includegraphics[width=9cm, height=7cm]{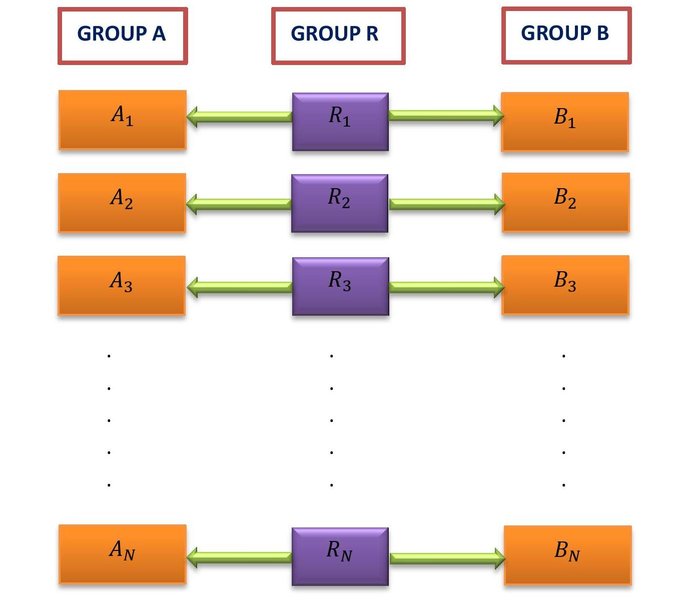}
\caption{\small $N$ pairs of players $\{A_i, B_i\}, i=1,2,3,...N$ are interrogated by $N$ referees. $R_i$ asks questions to $A_i$ and $B_i$. Groups $A$ and $B$ cooperate using a predetermined strategy with an aim to maximize their payoff.}
\end{figure}

For the ease of establishing correspondences between the questions and the measurement choices, and between the answers and the measurement outcomes, we choose $\mathcal{K}=\{1,2\}, \mathcal{L}=\{2,3\}$, and $\mathcal{R}, \mathcal{S}=\{0,1\}$. In the limit $N\rightarrow\infty$, we assume that we have well-defined normalized probability distributions $\xi(k,l)$ and $\alpha(r,s|k,l)$. We define the payoff function as follows:

\[ \mu(r,s|k,l) =
  \begin{cases}
    +1       & \quad \text{if }  r\oplus s=(kl)\bmod2\\
    -1  & \quad \text{otherwise } \\
  \end{cases}
\]

Then, the average payoff obtained by the groups A and B can be computed as $\expval{\mu(r,s|k,l)}_N$ where the averaging is carried out by considering the relevant payoffs secured by all of the $N$ pairs $\{A_i,B_i\}$, $i=1,2,...,N$.\\

Let us now presume that temporal correlations of measurements performed on the same system are manipulated by the two groups of players in the following manner: for the pair of questions $\{r,s\}$, $A_i$ and $B_i$ perform projective spin measurements on a shared quantum bit and notify their answers depending on the outcomes observed in the respective cases. Temporal correlations are defined by correlations between the observed outcomes of measurements effected on the same system at different instants in time. Here, we shall concern ourselves with two-point correlations arising from measurements  performed at two distinct times. As we have already made it clear, there is a bijective mapping between the outcomes and the responses, and therefore, we shall use $\{r, s\}$ to signify not only the pair of responses but also the pair of observed outcomes. So we invoke the following correspondence: $\mathcal{R},\mathcal{S} \rightarrow \{+1,-1\}$ (the set of observed outcomes). That is, the first element of either set corresponds to an observed outcome of $+1$ and the second to $-1$. Similarly, we use $k$ and $l$ to denote the answers as well as the observed outcomes. \\

For the questions $1$ and $2$ posed to $A_i$, he performs measurements of the same spin projection, say along $\hat{r}$, at times $t_1$ and $t_2$ respectively, and for the questions $2$ and $3$ made to $B_i$, the latter performs measurements along $\hat{r}$, at times $t_2$ and $t_3$ respectively, on a shared two-level quantum system. The measurement rules in this game are delineated in FIG. 3.\\

\begin{figure}[ht!]
\centering
\includegraphics[width=9cm, height=5cm]{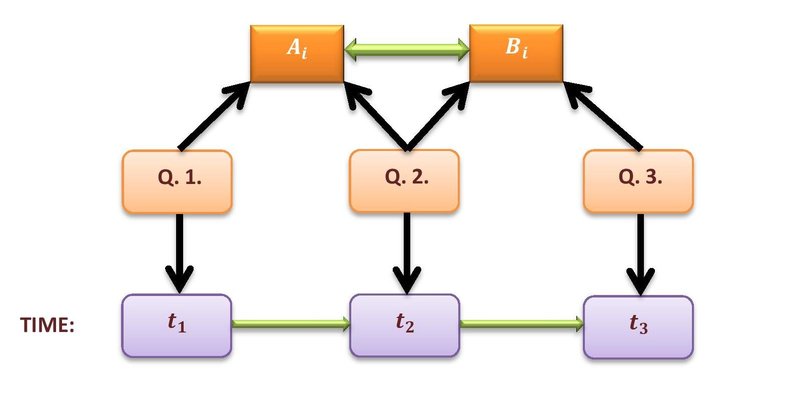}
\caption{\small $A_i$ can be asked one of the questions $Q. 1.$ and $Q. 2.$; $B_i$ can be asked one of the questions $Q. 2.$ and $Q. 3.$ . If $Q. 1.$ is asked, $A_i$ performs  measurement at $t_1$; if $Q. 2.$ is asked, $A_i/B_i$ performs a measurement at $t_2$; if $Q. 3.$ is asked, $B_i$ performs a measurement at $t_3$. $t_1 < t_2 < t_3$}
\end{figure}

It is easy to see that the temporal correlator $C_{kl}$ can now be expressed as:

\begin{align}
C_{kl}&=\sum_{r\oplus s=0}\alpha(r,s|k,l)-\sum_{r\oplus s=1}\alpha(r,s|k,l)
\end{align}
Defining the variable $t=r\oplus s$ and invoking the normalization condition $\sum_{r\oplus s=0}\alpha(r,s|k,l)+\sum_{r\oplus s=1}\alpha(r,s|k,l)=1$, we have the following compact relation:

\begin{align}
\sum_{r\oplus s=t}\alpha(r,s|k,l)=\frac{1}{2}(1+(-1)^tC_{kl}), t\in\{0,1\} \label{e6}
\end{align}

The probability distribution $\xi(k,l)$ is defined as follows:

\begin{align}
\xi(k,l)=\frac{1}{3}(1-\delta_{kl}) \label{e7}
\end{align}

The consequence of such a definition is that if one of the two players $A_i$ and $B_i$ is asked the question $2$ corresponding to his/her set of questions, the other will definitely not be asked the question $2$ from his/her set. The rest of the
question pairs, $\{1, 2)\}$, $\{1, 3\}$ and $\{2, 3\}$ are all equi-probable. Note that this gives either member of the pair $\{A_i , B_i\}$ the chance to unambiguously make out the question posed to his/her partner, if and when he/she is asked the question $2$ from his/her set. However, the players won't be knowing each other's individual responses and therefore, will not be able to decide upon a pair of answers that would fetch them the higher payoff of +1. They would thus be compelled to play the game probabilistically. Further, also note that since $\xi(2,2)=0$, there is no possibility of $A_i$ and $B_i$ making the same measurement at the same time.\\

Having stated all the relevant parameters, we define $\Pi_{LGI}=\lim_{N\rightarrow\infty}\expval{\mu}_N $ so that upon substitution of $\ref{e6}$ and $\ref{e7}$ in $\ref{e3}$ we obtain:
\begin{align}
\Pi_{LGI}&=\frac{1}{6}\sum_{k,l,t} (-1)^{\delta_{t,(kl)\bmod2}}(\delta_{kl}-1)(1+(-1)^tC_{kl})\\
&=\frac{1}{3}(C_{12}+C_{23}-C_{13})
\end{align}
where $C_{12}+C_{23}-C_{13}=\Delta_L$ resembles the well known expression corresponding to the Leggett-Garg inequality (LGI) $\Delta_L\leqslant 1$ (\cite{39}). The LGI is based on the assumptions of realism and noninvasive measurability, but holds true whenever the joint probabilities of outcomes obtained on measuring the same physical quantity at two different times are factorizable. However, if the temporal correlations are quantum (\cite{39},\cite{61},\cite{62}) in nature, where measurement invasiveness does not usually allow the probabilities of outcomes at two different times to be independent of one another, $\Delta_L$ can attain a highest value of $\dfrac{3}{2}$.\\

 It thus becomes evident in this backdrop that while ideal noninvasive (classical) measurements can yield an average payoff of at most $0.33$ (approx.), manipulation of invasive quantum measurements can lead to a maximum average payoff of $0.5$ (a staggering $51.51\%$ hike in the expected payoff). Alternatively speaking, if the game is played using resources in which there exists no correlation between temporally separated measurements performed on a system, the average payoff is bounded above by a numerical value of $0.33$. However, if we make use of quantum resources so that measurements performed are inherently invasive in nature, we can go beyond this classical upper constraint by about $51.51\%$.    Also $\Pi_{LGI}=0.5$ represents a quantum correlated Nash equilibrium as this is the highest possible payoff attainable by $A$ and $B$ using quantum resources and the players won't have any incentive to switch to an alternative strategy. 
\section{A Biased LGI Game} \label{s4}

We shall now continue to consider a scenario identical to the Bayesian LGI Game with a minor alteration of its rules. In place of the
group of $N$ referees, we consider a group $C$ of $N$ players $C_1,C_2,...,C_N$ not only performing the same task as the referees in the previous game but also actively participating in the game. However, we no more wish to treat $\mu(r,s|k,l)$ as the payoff function assigned to $A_i$ and $B_i$ for the pair of responses $\{r,s\}$ on being asked the pair of questions $\{k,l\}$. Instead, we define it as ``points earned by the pair $\{A_i, B_i\}$" under the same circumstances. \\

In this context, we construct the quantity $\zeta={\mu}-\mu_{cl}$, where $\mu_{cl}=\dfrac{\Delta_{L}^{cl}}{3}=0.33$ and $\Delta_{L}^{cl}=1$ corresponds to the upper bound for the Leggett-Garg Inequality. But in this new game, we insist on the active participation of not only the members of $A$ and $B$, but also of those belonging to the group $C$. If $A$ and $B$ win the game, the group $C$ loses, and vice-versa. Thus, on the one side, we have a cooperation between $A$ and $B$, both of whom have conflict of interests with the third group $C$. The following winning criterion is imposed for $A$ and $B$: 

\begin{align}
&\expval{\zeta}_{N} > 0\\
\implies &\expval{\mu} >\mu_{cl}
\end{align} 

That is, if the average number of points earned by $A$ and $B$ exceeds the classical upper limit of $0.33$, they win and $C$ loses the game.\\

 As we have observed in the original version of the game, (in the limit $N\rightarrow\infty$) $A$ and $B$ can make use of suitable temporal quantum correlations which lead to a violation of the Leggett-Garg inequality and can definitely win this game. We, therefore, consider the following question: is this game perfectly biased in favour of $A$ and $B$, so that the group $C$ is doomed to lose? The answer is an emphatic ``no". There, in fact, exists a nice strategy for $C$ that would shift the balance completely in its favour. \\

To this end, we elucidate an advantageous feature of \textit{intervening} measurements performed at some time(s) \textit{between} two temporally separated measurements, all of which are carried out on the same system. In the ordinary LGI game corresponding to a given run, $A_i$ is assumed to perform a measurement at time $t_A$ (say), following which, $B_i$ makes a measurement at a time $t_B (>t_A)$ (say) on the same system. But here we wish to investigate a situation where, over and above these measurements, one or more number of measurements are carried out on that system at some \textit{intermediate} times. We shall deal only with projective measurements for simplicity. A further simplification is attained by setting the Hamiltonian of the system, $\hat{H}$, to zero, so that the collapse of the state vector upon a measurement is the only dynamics present in the qubit. However, in that case, there is no such restriction placed on the directions of measurements effected by $A_i$ and $B_i$ (recall that in the original game we had imposed the condition that the same spin projection is measured at different times). In the Heisenberg picture, the observable $\hat{Q}$ being measured by $A_i$ and $B_i$ evolves unitarily from $t_A$ to $t_B$ in conformity with the rule

\begin{align}
\hat{Q}(t_B)=\exp(i\hat{H}(t_B-t_A))\hat{Q}(t_A)\exp(-i\hat{H}(t_B-t_A))
\end{align}

But setting $\hat{H}=0$ causes the state to be frozen in time, so, in the Heisenberg picture, the observables don't evolve either. As a result,  we have $\hat{Q}(t_3)=\hat{Q}(t_2)=\hat{Q}(t_3)$. That is why, we relax the restriction of measuring the same observable in this context and allow for the measurement of three observables corresponding to three arbitrarily chosen directions of spin.\\

 Assume now that a total of $n$ intermediate spin measurements are executed on the given system at times $t_I^1, t_I^2,...,t_I^n$ along the directions $\hat{r}_1,\hat{r}_2,...,\hat{r}_n$ respectively, where $t_A<t_I^1<t_I^2<t_I^2<...<t_I^n<t_B$. Further suppose that $A_i$ and $B_i$ perform their spin measurements along $\hat{a}$ and $\hat{b}$ respectively and denote the intermediate measurement outcomes, in chronological order, by $x_1,x_2,...,x_n$. Under such a condition, it can be shown (see Appendix \ref{II}) that we are led to the following expression for the temporal correlator between the outcomes obtained by $A$ and $B$:

\begin{align}
C'_{\hat{a},\hat{b}}=\beta_n(\hat{a}\vdot\hat{r}_1)(\hat{r}_n\vdot\hat{b}) \label{e13}
\end{align} 
where $\beta_n=(\hat{r}_1\vdot\hat{r}_2)(\hat{r}_2\vdot\hat{r}_3)...(\hat{r}_{n-1}\vdot\hat{r}_n)$. For only one intermediate measurement performed in the direction $\hat{c}$, the following form can be inferred:
\begin{align}
C'_{\hat{a},\hat{b}}=(\hat{a}\vdot\hat{c})(\hat{c}\vdot\hat{b}) \label{e14}
\end{align}

Herein we wish to manifest a novel feature of correlators of the above form. A straightforward substitution into the third-order Leggett-Garg expression reveals that the correlators defined in $\ref{e14}$ are incapable of violating the LGI, i.e. $\Delta_L\leqslant1$. In other words, an intervening projective measurement corresponding to each of the sub-ensembles defining $C_{12}, C_{23}$ and $C_{13}$ ensures that the Leggett-Garg inequality is satisfied and the invasiveness of the measurement performed by Alice ``apparently" disappears.

\begin{align}
(\hat{a}_1\vdot\hat{c})(\hat{c}\vdot\hat{a}_2)+(\hat{a}_2\vdot\hat{c})(\hat{c}\vdot\hat{a}_3)-(\hat{a}_1\vdot\hat{c})(\hat{c}\vdot\hat{a}_3) \leqslant 1 \label{e15}
\end{align}

Note that we have explicitly assumed here that the same projective measurement (along $\hat{c}$) is carried out at some intermediate time as regards to each of the three sub-ensembles. Equation \ref{e15} is easy to verify, however for a simple proof, one is referred to Appendix \ref{II}.\\

\begin{figure}[ht!]
\centering
\includegraphics[width=8cm, height=4cm]{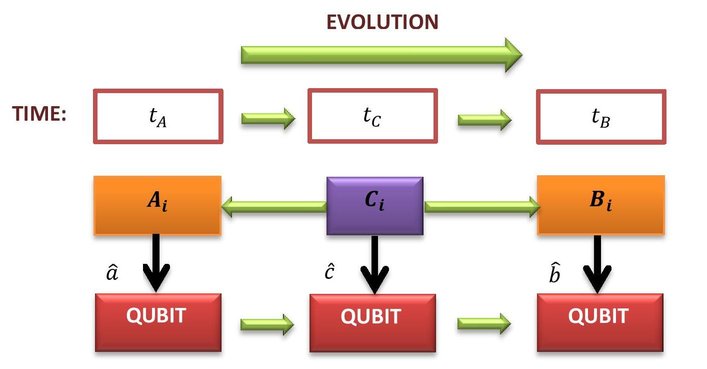}
\caption{\small $A_i$ performs a spin measurement along $\hat{a}$ at $t_A$; $B_i$ measures along $\hat{b}$ at $t_B$. $C_i$ intervenes at an intermediate time $t_C$ performing a measurement along $\hat{c}$. This intervention modifies the correlations between the measurement outcomes of $A_i$ and $B_i$ according to equation \ref{e14} . $t_A<t_C<t_B$}
\end{figure}

The apparent reversion to classicality as a consequence of an intervening measurement is precisely what could be exploited in this game if the group $C$ aspires to shift the bias of the game in its favour. As is clearly discernible at this juncture, the members of $C$ could conspire, in unison, to induce the \textit{same} projective measurement on the system shared by each of the pairs $\{A_i, B_i\}$ so as to guarantee that any advantage gained by the groups $A$ and $B$ over the classical upper bound for the average number of points earned vanishes. In this case, since $\Delta_L\leqslant 1$, or equivalently $\expval{\mu}\leqslant\expval{\mu_{cl}}$, $A$ and $B$ would never be able to win the game, no matter what spin measurements each of the pairs $\{A_i, B_i\}$ carries out on the shared system. We have thus illustrated the existence of a simple but useful strategy available to $C$ that could effectively rob $A, B$ of the counterfactual super-classical advantage that would have been achieved had there been no intervention. In fact, it doesn't matter either as to what intermediate measurement is performed by each of the referees; equation $\ref{e15}$ holds for any unit vector $\hat{c}$. Therefore, as long as $C_i$ performs a projective spin measurement in between $t_A$ and $t_B$ on the system shared by $A_i$ and $B_i$, it is guaranteed that $C$ will outsmart $A$ and $B$.

\section{A Completely Biased Temporal CHSH Game} \label{s5}
This game too has a similar setting as the LGI one (\ref{s3}), albeit with a few essential modifications. In this case, we choose all of the sets $\mathcal{K}, \mathcal{L}, \mathcal{R}$ and $\mathcal{S}$ to be $\{0,1\}$ and continue to denote, by $\mu(r,s|k,l)$, the number of points earned by $A_i$ and $B_i$ when they provide the pair of responses $\{r, s\}$ to the pair of questions $\{k, l\}$ asked by $C_i$. However, the function $\mu$ is also modified (\cite{18}), as delineated below:

\[ \mu(r,s|k,l) =
  \begin{cases}
    +1       & \quad \text{if }  r\oplus s=kl\\
    -1  & \quad \text{otherwise } \\
  \end{cases}
\]

The distribution function corresponding to any pair of questions is taken to be uniform:
\begin{align}
\xi(k,l)=\frac{1}{4} \forall k\in\mathcal{K},l\in\mathcal{L}
\end{align}

The rules remain the same as the Biased LGI Game, apart from the fact that $\mu_{cl}$ is now equal to $\dfrac{\Delta_{CHSH}^{cl}}{4}=0.5$, where $\Delta_{CHSH}^{cl}=2$ is the upper bound of the temporal CHSH inequality (\cite{50},\cite{60},\cite{61}):
\begin{align*}
\Delta_{CHSH}=\expval{Q_0(t_A)R_0(t_B)}+\expval{Q_0(t_A)R_1(t_B)}+\\\expval{Q_1(t_A)R_0(t_B)}-\expval{Q_1(t_A)R_1(t_B)}\\ \leqslant 2
\end{align*}

 In the temporal CHSH scenario, there are two parties with the respective measurement choices $\{Q_0, Q_1\}$ and $\{R_0, R_1\}$. The first party measures one variable from the set $\{Q_0, Q_1\}$ at an earlier time $t_A$ and the second party measures another from the set $\{R_0, R_1\}$ at a later time $t_B$, and correlators corresponding to each of the $4$ pairs of measurements, ${Q_i,R_j}$, $i,j \in\{0,1\}$, are estimated. All variables are assumed to be dichotomic, i.e. $Q_i,R_j=\pm1$ and measurements pertain to a single qubit.  \\

In the context of this game, $A_i$ is assumed to perform a projective measurement at $t_A$ and $B_i$ another projective measurement at $t_B (>t_A)$ on the same system. The measurement rules are outlined in the following table:\\
\begin{table}[ht]
\centering
\begin{tabular}{||c |c |c||}
\hline 
\bf{Q. No.} & \bf{S. M. by $A_i$} & \bf{S. M. by $B_i$}\\
\hline
$0$ & along $\hat{a}_0$; T. M. = $t_A$ & along $\hat{b}_0$; T. M. = $t_B$ \\
\hline
$1$ & along $\hat{a}_1$; T. M. = $t_A$ & along $\hat{b}_1$; T. M. = $t_B$ \\
\hline
\end{tabular}\\
\textbf{S. M.} $\rightarrow$ Spin Measurement\\
\textbf{T. M.} $\rightarrow$ Time of Measurement
\end{table}

If the temporal correlations between the measurement outcomes obtained by $A_i$ and $B_i$ are utilized, then in view of the fact that the relation $\sum_{r\oplus s=t}\alpha(r,s|k,l)=\frac{1}{2}(1+(-1)^tC_{kl})$ continues to hold as we persist with the mappings $\mathcal{K},\mathcal{L}\rightarrow\{+1,-1\}$, we uncover the following relation for $\Pi_{CHSH}=\lim_{N\rightarrow\infty}\expval{\mu}_N$:

\begin{align}
\Pi_{CHSH}&=\frac{1}{4}\sum_{k,l,t}(-1)^{\delta_{t,kl}}(1+(-1)^tC_{kl})\\
&=\frac{\Delta_{CHSH}}{4}
\end{align} 
and accordingly, an upper bound for ideal classical measurements is obtained :
\begin{align}
\Pi_{CHSH}\leqslant \Pi_{CHSH}^{cl}=0.5
\end{align}

\begin{figure}[ht!]
\centering
\includegraphics[width=9cm, height=4cm]{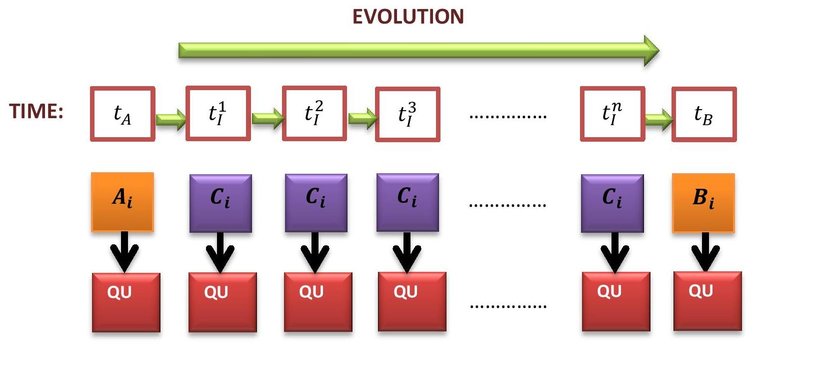}
\caption{\small $A_i$ and $B_i$ perform spin measurements on the qubit at $t_A$ and $t_B$ respectively. $C_i$ performs multiple interventions at intermediate times $t_I^1, t_I^2,...t_I^n$ on the same qubit that is shared by $A_i$ and $B_i$. Quite remarkably, for arbitrary number of interventions, the temporal CHSH inequality pertaining to the correlations between the measurements of $A_i$ and $B_i$ is always satisfied.}
\end{figure}

But suitable quantum temporal correlations will result in a maximum value of about 0.707 ($\approx 41.42\%$ increase) for $\Pi_{CHSH}$, as the Tsirelson's Bound (\cite{50},\cite{64}) for the temporal CHSH inequality happens to be $2\sqrt{2}$ (identical to the spatial scenario (\cite{64})). Therefore, as per the rules of the game, whenever the members of $A$ and $B$ operate with correlations that precipitate in a violation of the temporal CHSH inequality, they win the game and $C$ ends up on the losing side. However, recall that in the biased LGI game (\ref{s4}), we proposed a novel strategy for the group $C$ enabling it to wrest out a win with certainty. Could that strategy be employed in this context as well? Indeed, it can. We observe, once again, that by performing an intermediate measurement it is possible to quell the super-classical advantage of the correlations exploited by $A$ and $B$. Not only that, the strategy in this case is found to extremely robust, as we shall soon be able to acknowledge. If we compute the CHSH expression corresponding to the temporal correlators (\ref{e13}) pertaining to the situation where arbitrary number of intermediate projective measurements are performed on the shared qubit, we find that it is also bounded above by $2$. 

\begin{align*}
\beta_n[{(\hat{a}_0\vdot\hat{r}_1)}{(\hat{r}_n\vdot\hat{b}_0)}+{(\hat{a}_0\vdot\hat{r}_1)}{(\hat{r}_n\vdot\hat{b}_1)} \\ +{(\hat{a}_1\vdot\hat{r}_1)}{(\hat{r}_n\vdot\hat{b}_0)}-{(\hat{a}_1\vdot\hat{r}_1)}{(\hat{r}_n\vdot\hat{b}_1)}]\leqslant 2
\end{align*}

Thus, quite remarkably, even when multiple intervening measurements are executed at different intermediate times on the system, the temporal CHSH inequality is still satisfied. This follows directly from the fact that $\beta_n$ is bounded above by 1, for all positive integral values of $n$. Consequently, it can be made out quite clearly that following any intermediate measurement executed by $C_i$, $B_i$ won't be able to retrieve any super-classical advantage by performing any additional (intermediate) measurement(s). Thus, the balance tilts irrevocably in the favour of $C$ precluding any possibility of $A$ and $B$ winning the game. \\

\section{Temporal Correlations in a Nonlocal Bayesian Game} \label{s6}
We have already mentioned before that the connection between Bayesian games and Bell Nonlocality was made explicit in the seminal work of Brunner and Linden (\cite{18}), in which the CHSH inequality (\cite{37}) was viewed through the lens of Bayesian games. Here we propose a slender variation of the nonlocal CHSH game by casting it in a similar setting as the temporal CHSH one. Plus, we extend the number of participants to three: the active participants are Alice, Bob and Charlie with the first two cooperating with each other against the third. Charlie asks questions to Alice and Bob following a distribution $\xi(k,l)$. The functions $\mu(r,s|k,l)$ and $\xi(k,l)$ are identical to those discussed in the biased temporal CHSH game (\ref{s5}):
\[ \mu(r,s|k,l) =
  \begin{cases}
    +1       & \quad \text{if }  r\oplus s=kl\\
    -1  & \quad \text{otherwise } \\
  \end{cases}
\]
\begin{align}
\xi(k,l)=\frac{1}{4} \forall k\in\mathcal{K},l\in\mathcal{L}
\end{align}

The winning condition for Alice and Bob continues to be dictated by $\expval{\zeta} > 0$, where the average is calculated from a sample of sufficiently large number of questions and the corresponding answers given in response. Alice and Bob share two suitably entangled qubits for which the Bell-CHSH inequality is violated. Alice has access to one of the qubits and Bob to the other. Each player performs measurement on the qubit that he/she can access. The measurement rules too remain the same as
those stated in the biased temporal CHSH game with one significant difference: Alice and Bob are hereby supposed to carry out measurements on different qubits and not on the same qubit. But like in the previous case, Alice always performs her measurement at an earlier time $t_A$ while Bob at $t_B(>t_A)$. An interesting feature of this game is that the correlations so generated between the measurement outcomes obtained by Alice and Bob are not merely ``nonlocal", but also ``temporal" in nature since there happens to be a ``temporal separation" between the two measurements. \\

As one can easily verify for oneself, because these measurements involve commuting observables (Alice and Bob perform local measurements on their qubits), the introduction of a temporal separation does not alter the inherently nonlocal correlations between the two qubits - in fact, the correlations behave as if both measurements are simultaneous (see Appendix \ref{IV}). That is, the correlator between $(\hat{Q}\otimes\mathds{1})$ measured at $t$ and $(\mathds{1}\otimes\hat{R})$ measured at $t+\Delta t$ is independent of the magnitude of $\Delta t$. Then $\Pi_{CHSH}^{nonlocal}=\dfrac{\Delta_{CHSH}^{nonlocal}}{4}$,
as before, regardless of the magnitude of the temporal separation.\\

However, because of the temporal separation, a novel prospect arises in the biased version of the game, where the winning criterion is identified with
the inequality $\expval{\zeta} > 0$. Once again, it is revealed that by incorporating intermediate measurements, the nonlocal correlations can be made to abide by the CHSH inequality, provided the intervention is made on Bob's qubit. In
that case, the correlators take the following form (refer to Appendix \ref{IV} for a derivation):
\begin{align}
C''_{\hat{a},\hat{b}}=\kappa_n C_{a,\hat{r}_1}
\end{align}

\begin{figure}[ht!]
\centering
\includegraphics[width=9cm, height=6cm]{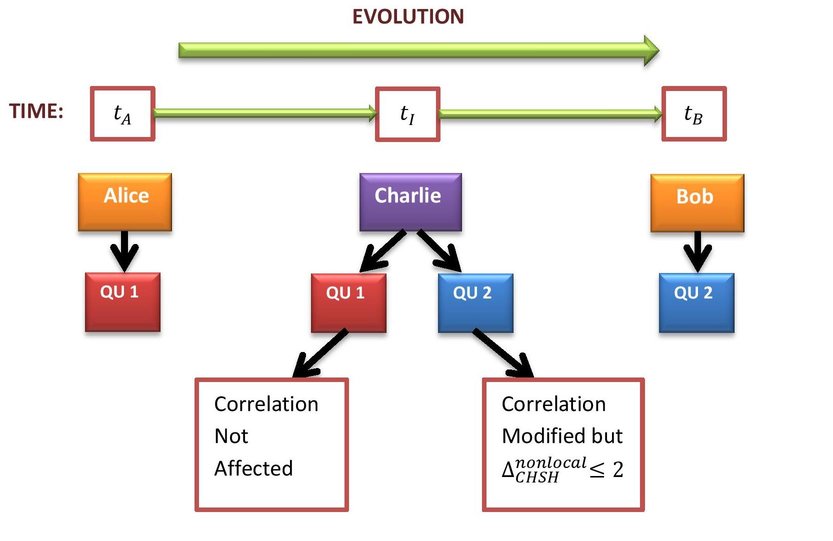}
\caption{\small Alice has the qubit QU1 in her possession and Bob has access to QU2. The two qubits are suitably entangled. Alice measures on QU1 at $t_A$ and Bob measures on QU2 at $t_B$. Charlie intervenes at an intermediate time $t_I$ on one of the two qubits. If he intervenes on Alice's qubit, the correlations between Alice and Bob's measurements are unaltered. However, if the intervention is made on Bob's qubit, the correlations are changed, albeit in a suitable manner so as to respect the Bell-CHSH inequality.}
\end{figure}

where $C''_{\hat{a},\hat{b}}$ refers to the correlation between the two outcomes when $n$ number of intervening measurements are carried out on Bob's system, $C_{\hat{a}, \hat{r}_1}$ to the correlation between the outcomes obtained by Alice and the outcomes due to the first intervening measurement performed on Bob's system, and $\kappa_n=(\hat{r}_1\vdot\hat{r}_2)(\hat{r}_2\vdot\hat{r}_3)...(\hat{r}_{n-1}\vdot\hat{r}_n)(\hat{r}_n\vdot\hat{b})$. Note that $\kappa_n$ is related to $\beta_n$ defined earlier, as 
\begin{align}
\kappa_n=(\hat{r}_n\vdot\hat{b})\beta_n
\end{align}

 And again, like in the temporal CHSH scenario, $\Delta_{CHSH}^{nonlocal}$ is found to be bounded above by $2$ (see Appendix \ref{IV})for an arbitrary number of intervening measurements - the scope of gaining super-classical advantage on part of Alice and Bob is perennially lost making it a completely biased game.
\begin{align*}
\beta_n[(\hat{b}_0\vdot\hat{c}+\hat{b}_1\vdot\hat{c})C_{\hat{a}_0,\hat{c}}\\+(\hat{b}_0\vdot\hat{c}-\hat{b}_1\vdot\hat{c})C_{\hat{a}_1,\hat{c}}]\leqslant 2
\end{align*}

 One should understand that this feature would have been impossible had there been no temporal separation between the measurements performed by Alice and Bob. It is only through this fact that we can realize the true essence of integrating temporal correlations in our game.\\

As a closing remark, we would like to comment on the case in which interventions are made on Alice's qubit and emphasize on the fact that in this scenario, the correlators between Alice and Bob's measurement outcomes remain invariant in form (see Appendix \ref{IV} for a brief discussion on this). Thus, it is absolutely necessary for Charlie to effect an intervening measurement only on Bob's qubit in between the measurements performed by Alice and Bob. He gains nothing by performing a measurement on Alice's qubit.

\section{Epilogue}
To put things into perspective, the essence of our endeavour has been to look beyond the much-discussed relevance and utility of nonlocal correlations for information processing and game-theoretic implementations. Quantum theory is rife with spooky, counter-intuitive features, and nonlocality, or more precisely, the conflict with a local realistic world view, is just one of them. Though admittedly, from the point of view of quantum foundations and its applications, the existence of nonlocality has usually captured the imagination of quantum physicists and philosophers more than anything else, there has been a plethora of literature centering this phenomenon. In particular, an overwhelming majority of information processing tasks and quantum games bank on the potential of nonlocal resources. Given this state of affairs, our exploration was driven by the motivation of presenting the applicability of temporal correlations realizable in the quantum realm, albeit in a fairly elementary form, to the restricted context of Bayesian games.\\

Game theory (\cite{24},\cite{25}) as a mathematical device appertains to the analytical  treatment of the decision-making strategies available, in principle, to the active participants of a game. Looking at correlation inequalities in quantum theory through the lens of games enables us to uncover innovative correlated strategies leading to interesting outcomes when the games are played using properly synchronized quantum devices. In the cooperative nonlocal Bayesian game (\cite{18}), the players share nonlocal resources, manipulation of which leads to correlated strategies that cannot be replicated using local resources and can successfully outperform all available classical strategies that rely on local resources. Consequently, these strategies make it possible for the participating parties to earn payoffs higher than the classical upper bound. In the Bayesian LGI game (\ref{s3}), appropriately correlated strategies are realized through temporal correlations that engender in a violation of the Leggett-Garg Inequality (\ref{e1}). These correlations stem from the property of measurement-invasiveness observed in quantum systems. Rigorously stated, invasiveness of measurements is a categorically nonclassical feature; while measurement on a classical system can be carried out by effecting an arbitrarily small perturbation on the ontological state of the system, in the quantum regime, a measurement will almost always lead to a disturbance in the ontic state, assuming, of course, that there exists an underlying reality to any quantum system.\\  

In sections \ref{s4} and \ref{s5} we have highlighted an interesting ``disentangling" effect of intermediate projective measurements carried out on a two-level quantum system between two projective measurements separated in time. In the Leggett-Garg scenario (\ref{s4}), where we consider the expression $\Delta_L=\expval{Q(t_1)Q(t_2)}+\expval{Q(t_2)Q(t_3)}-\expval{Q(t_1)Q(t_3)}$, we observe that under the insertion of an intervening projective measurement, any possibility of a violation of the LGI is done away with. This provides the group $C$ in our Biased LGI Game with a ``quantum" strategy to successfully thwart the cooperation of $A$ and $B$. That this contrivance on part of $C$ is being branded ``quantum" is consistent with the fact that resources used in our game are unmistakably quantum and so are all the measurement schemes. Consequently, this semblance of a reversion to classicality is a phenomenon noticed exclusively within the quantum paradigm without taking recourse to any classical devices. The groups $A$ and $B$ are obviously not interested in utilizing classical correlations; it is the group $C$ that has a conflict of interest with the former and is therefore, intent upon curbing any super-classical advantage that $A$ and $B$ might get out of the quantum resources in their possession.\\

 It is further demonstrated in section \ref{s5} that the temporal correlators are modified, as they should be, in such a fashion as to suppress any (counterfactual) violation of the temporal CHSH inequality ($\Delta_{CHSH}=\expval{Q_0(t_A)R_0(t_B)}+\expval{Q_0(t_A)R_1(t_B)}+\expval{Q_1(t_A)R_0(t_B)}-\expval{Q_1(t_A)R_1(t_B)}\leqslant 2$). What's more, in this scenario, the correlation inequality is satisfied for any number of intermediate measurements. This provides $C$ the possibility of winning the biased temporal CHSH game unconditionally, as once an intervention is made, the shared system cannot be manipulated any further to generate correlations capable of violating the relevant inequality. \\
 
 Finally, in section \ref{s6}, we introduce a temporal separation in the customary EPR-Bohm type setting in which, once again, intermediate intervention provides the novel possibility of reverting to correlations that can be simulated by local correlations. This is because any correlation that respects the spatial CHSH inequality (\cite{37}) can be mimicked using appropriately chosen local correlations. This possibility is made available only in a hybrid spatio-temporal setting where there is a time gap between the measurements of Alice and Bob (and therefore, between the times when the two cooperators are asked their respective questions).\\

We conclude with a few remarks in relation to our attempt to exemplify some of the applicative facets of temporal quantum correlations. We have chosen to deal explicitly with the restricted class of sharp projective measurements effected on two-level quantum systems, obviously for the sake of mathematical convenience. Consequently, we leave open the possibility of exploring the corresponding correlations when the measurement schemes are different - as for example, when unsharp measurements, POVM, etc. are employed, and/or when higher dimensional systems are utilized. One could also consider the potential applications of temporal correlations in information processing tasks, namely  in the context of cryptography, quantum communication and communication complexity. In such scenarios, it could be worth investigating the relevance of an intervening measurement. An intervention could possibly be construed as an attack on effective quantumness or an act of impeding quantum advantage, as has been demonstrated in the games we formulated. However, if such interventions could enhance quantum advantage in certain game-theoretic settings or information processing tasks may also be probed. 

\section*{Acknowledgements}
The author is grateful to Dipankar Home and Som Kanjilal for fruitful discussions and for their insightful suggestions in course of this work.

\section*{Appendix}

\subsection*{I} \label{I}
A crucial result was used in the context of the biased Bayesian LGI game (\ref{s4}) that when a single intervening projective measurement is carried out on the system shared by $A_i$ and $B_i$, the resulting correlations between the outcomes obtained by $A$ and $B$ are rendered incapable of violating the Leggett-Garg Inequality (\ref{e1}). In the context of the Biased Temporal CHSH game (\ref{s5}), this feature was claimed to be holding true for any number of interventions. Before we can prove this result, we need to work out an expression for the temporal correlator between the outcomes obtained by two parties. We begin by reiterating the following assumptions:\\

(i) A total of $n$ intermediate spin measurements are executed on the given system at times $t_I^1, t_I^2,...,t_I^n$ along the directions $\hat{r}_1,\hat{r}_2,...,\hat{r}_n$ respectively, where $t_A<t_I^1<t_I^2<t_I^2<...<t_I^n<t_B$. \\

(ii) $A_i$ and $B_i$ perform their spin measurements along $\hat{r}_0$ and $\hat{r}_{n+1}$ respectively getting outcomes $x_0$ and $x_{n+1}$ respectively. (We have tweaked the notations a bit for our convenience.) We denote the intermediate measurement outcomes, in chronological order, by $x_1,x_2,...,x_n$. \\

The temporal correlator between the outcomes obtained by $A_i$ and $B_i$ are calculated as:

\begin{align*}
C_{\hat{r}_0,\hat{r}_{n+1}}=\sum_{x_0,x_1,x_2,...,x_n,x_{n+1}=\pm1} x_0x_{n+1}.\prod_{i=0}^{n-1}\Tr(\chi_{r_{i-1}}^{x_{i-1}}\chi_{r_{i}}^{x_{i}}) \label{9}
\end{align*}  

Here, $\chi_{r_i}^{x_i}=\dfrac{1}{2}(\mathds{1}+x_i\va{\sigma}\vdot\hat{r}_i)$, $\chi_{r_{-1}}^{x_{-1}}=\rho=\dfrac{1}{2}(\mathds{1}+\va{\sigma}\vdot\hat{r}_{-1})$ (the initial state of the system) and $\mathds{1}$ is the two-dimensional Identity operator. To begin simplifying this problem, we first take into account the last two Trace-terms in 
\begin{align*}
\prod_{i=0}^{n-1}\Tr(\chi_{r_{i-1}}^{x_{i-1}}\chi_{r_{i}}^{x_{i}})
\end{align*}

and carry out the summation over the variable $x_{n+1}$. Then, in view of the fact that 
\begin{align*}
\sum_{x_{n+1}=\pm 1}x_{n+1}\chi_{r_{n+1}}^{x_{n+1}}
\end{align*}

 is simply the projector-decomposition of the spin observable along $\hat{r}_{n+1}$, i.e. $(\va{\sigma}\vdot\hat{r}_{n+1})$, we have:
\begin{align*}
\sum_{s=\pm1}\Tr(\chi_{r_n}^{x_n}\chi_{r_{n+1}}^{x_{n+1}})&=\frac{1}{2}\Tr((\mathds{1}+x_n(\va{\sigma}\vdot\hat{r}_n))(\va{\sigma}\vdot\hat{r}_{n+1}))\\
&=x_n(\hat{r}_n\vdot\hat{r}_{n+1})
\end{align*}
Here we have made use of the fact that $(\va{\sigma}\vdot\hat{a}_1)(\va{\sigma}\vdot\hat{a}_2)=(\hat{a}_1\vdot\hat{a}_2)\mathds{1}+i\va{\sigma}\vdot(\hat{a}_1\cross\hat{a}_2)$. Next we carry out the summation over $x_n$ so that another partially simplified expression is obtained:
\begin{align*}
\sum_{x_n,x_{n+1}=\pm1}\Tr(\chi_{r_{n-1}}^{x_{n-1}}\chi_{r_n}^{x_n})\Tr(\chi_{r_n}^{x_n}\chi_{r_{n+1}}^{x_{n+1}})\\
=\sum_{x_n=\pm1}x_n(\hat{r}_n\vdot\hat{r}_{n+1})\Tr(\chi_{r_{n-1}}^{x_{n-1}}\chi_{r_n}^{x_n})\\
=\frac{1}{2}(\hat{r}_n\vdot\hat{r}_{n+1})\Tr((\mathds{1}+x_{n-1}(\va{\sigma}\vdot\hat{r}_{n-1}))(\va{\sigma}\vdot\hat{r}_n))\\
=x_{n-1}(\hat{r}_{n-1}\vdot\hat{r}_n)(\hat{r}_n\vdot\hat{r}_{n+1})
\end{align*}
Continuing in this fashion, we find, at a certain stage:
\begin{align*}
C_{\hat{r}_0,\hat{r}_{n+1}}&={\beta_n}{(\hat{r}_n\vdot\hat{r}_{n+1})}\sum_{x_0,x_1=\pm1} x_0x_1\Tr(\rho\chi_{r_0}^{x_0})\Tr(\chi_{r_0}^{x_0}\chi_{r_1}^{x_1})\\
&={\beta_n}{(\hat{r}_n\vdot\hat{r}_{n+1})}\sum_{x_0=\pm1}x_0\Tr(\rho\chi_{r_0}^{x_0})\Tr(\chi_{r_0}^{x_0}(\va{\sigma}\vdot\hat{r}_1))\\
&={\beta_n}{(\hat{r}_n\vdot\hat{r}_{n+1})}\sum_{x_0=\pm1} \Tr(\rho\chi_{r_0}^{x_0})(\hat{r}_0\vdot\hat{r}_1)\\
&=\beta_n{(\hat{r}_0\vdot\hat{r}_1)}{(\hat{r}_n\vdot\hat{r}_{n+1})}\Tr(\rho\sum_{x_0=\pm1}\chi_{r_0}^{x_0})\\
&=\beta_n{(\hat{r}_0\vdot\hat{r}_1)}{(\hat{r}_n\vdot\hat{r}_{n+1})}\Tr(\rho)\\
&=\beta_n{(\hat{r}_0\vdot\hat{r}_1)}{(\hat{r}_n\vdot\hat{r}_{n+1})}
\end{align*}
This is identical to the expression in \ref{e14} with $\hat{r}_0=\hat{a}$, $\hat{r}_{n+1}=\hat{b}$. Recall that we had defined the quantity $\beta_n$ as being equal to $(\hat{r}_1\vdot\hat{r}_2)(\hat{r}_2\vdot\hat{r}_3)...(\hat{r}_{n-1}\vdot\hat{r}_n)$ and so, $\abs{\beta_n}\leqslant 1$. If only one intervention is made, it is easily seen that the correlator-expression is given by:
\begin{align*}
C_{\hat{r}_0,\hat{r}_2}^{(1)}={(\hat{r}_0\vdot\hat{r}_1)}{(\hat{r}_1\vdot\hat{r}_2)}
\end{align*}
where the direction of intermediate spin measurement is along $\hat{r}_1$. This particular expression (for the case when only one intermediate measurement is performed on the system) was stated by Brukner et al (\cite{50}). Here we have derived a generalized simplified expression for the temporal correlator when an arbitrary number of projective measurements are carried out between the the ones effected by $A_i$ and $B_i$.\\

 The following section justifies why the correlators can never violate the Leggett-Garg Inequality, when only one intervention is made on the qubit.

\subsection*{II} \label{II}
As we have just derived, the temporal correlator under consideration is suitably modified when an intermediate projective spin measurement is performed on the system, so that we have the following form:

\begin{align*}
C_{\hat{r}_0,\hat{r}_2}^{(1)}={(\hat{r}_0\vdot\hat{r}_1)}{(\hat{r}_1\vdot\hat{r}_2)}
\end{align*}

Let us now investigate the resulting LGI expression $\Delta_L=C_{12}+C_{23}-C_{31}=(\hat{a}_1\vdot\hat{r}_1)(\hat{r}_1\vdot\hat{a}_2)+(\hat{a}_2\vdot\hat{r}_1)(\hat{r}_1\vdot\hat{a}_3)-(\hat{a}_1\vdot\hat{r}_1)(\hat{r}_1\vdot\hat{a}_3)$. Taking $(\hat{a}_i\vdot\hat{r}_1)=\cos\phi_i$, $i=1,2,3$, we have:
\begin{align*}
\Delta_L &=[\cos\phi_1\cos\phi_2+\cos\phi_2\cos\phi_3-\cos\phi_1\cos\phi_3]\\
&=\cos\phi_2(\cos\phi_1+\cos\phi_3)-\cos\phi_1\cos\phi_3
\end{align*}
Then, clearly $\Delta_L\leqslant\Gamma=\max(\mathcal{E}_1,\mathcal{E}_2)$, where $\mathcal{E}_1=\cos\phi_1+\cos\phi_3-\cos\phi_1\cos\phi_3$, and $\mathcal{E}_2=-\cos\phi_1-\cos\phi_3-\cos\phi_1\cos\phi_3$. Consequently, if we can show that $\Gamma$ is never greater than $1$ (which is the classical upper bound for the Leggett-Garg Inequality), then basically we are done.\\

\textbf{Case 1: $\Gamma =\mathcal{E}_1$.} So we probe the upper bound of $\mathcal{E}_1$. Note that the expression $\mathcal{E}_1(\cos\phi_1,\cos\phi_3)=\cos\phi_1(1-\cos\phi_3)+\cos\phi_3$ is monotonically increasing in $\cos\phi_1$, since  $\cos\phi_3\leqslant1$. 
\begin{align*}
\therefore \max_{\phi_1}\mathcal{E}_1=(1-\cos\phi_3)+\cos\phi_3=1
\end{align*}

\textbf{Case 2: $\Gamma =\mathcal{E}_2$.} We now consider the upper bound of $\mathcal{E}_2$. In this case, the expression $\mathcal{E}_2(\cos\phi_1,\cos\phi_3)=-\cos\phi_1(1+\cos\phi_3)-\cos\phi_3$ is monotonically decreasing in $\cos\phi_1$, as a result of which, 
\begin{align*}
\max_{\phi_1}\mathcal{E}_2=(1+\cos\phi_3)-\cos\phi_3=1
\end{align*}

Since in either case it is clear that $\max_{\phi_1,\phi_2} \Delta_L=1$, our claim is vindicated.

\subsection*{III}

In this section we show that the temporal correlators between the observed outcomes of $A$ and $B$ as a consequence of an arbitrary number of intermediate measurements on the shared systems are found to abide by the temporal CHSH inequality. But before that, we briefly recapitulate the temporal version (\cite{50},\cite{60},\cite{61}) of the CHSH inequality.\\

The temporal CHSH inequality is usually derived from two premises: (a) \textit{Realism (R)}  $\rightarrow$ The results of measurement performed on a given physical system are determined by well-defined pre-existing properties pertaining to the system that exist independent of observation, and (b) \textit{Non-invasive Measureability (NIM)} $\rightarrow$ An ideal measurement does not perturb the ontic state of the system under consideration. Stated somewhat differently, it is possible, in principle, to extract all relevant information about the state of the system with arbitrarily small disturbance on its subsequent dynamics. Strictly speaking, a third assumption also goes into the derivation: \textit{Induction}. Essentially this means that the measurement outcome at any point in time does not depend on what is measured on the system at some other instant of time.\\

Consider a quantum system described by an underlying Hilbert space $\mathcal{H}$ and dynamics governed by the Hamiltonian $\mathcal{H}$. We have two sets of dichotomic observables $\{\hat{Q}_0,\hat{Q}_1\}$ and $\{\hat{R}_0,\hat{R}_1\}$, each with eigenvalues $\pm1$. Alice has the choice of measuring one out of $\hat{Q}_0, \hat{Q}_1$ at time $t_A$ and Bob can measure any one out of $\hat{R}_0, \hat{R}_1$ at time $t_B (>t_A)$. Then the temporal CHSH expression $\Delta_{CHSH}$ is given by:
\begin{align*}
\Delta_{CHSH} &=C_{00}+C_{01}+C_{10}-C_{11}\\
\end{align*}
where $C_{ij}=\dfrac{1}{2}\expval{\{\hat{Q}_i,\hat{R}_j\}}$ (\cite{60}) is the temporal correlator of the outcomes $Q_i\in\{+1,-1\}$ and $R_j\in\{+1,-1\}$ obtained on measuring the observables $\hat{Q}_i$ and $\hat{R}_j$ at times $t_A$ and $t_B$ respectively. Now, starting from the assumptions of \textit{Realism} and \textit{Noninvasive Measureability}, we shall show that $\Delta_{CHSH}\leqslant2$. To that end, we define ontic states $\lambda$ (\cite{54}-\cite{58},\cite{61}) with $\mu(\lambda)$ being the probability (density) that the system is prepared in the ontic state $\lambda$. So $\int d\lambda\mu(\lambda)=1$. Let $\xi_{i}(Q_i|\lambda)$ be the outcome function that signifies the probability of obtaining $Q_i$ on measuring the system in the ontic state $\lambda$ at time $t_i$. Similarly, let $\xi'_{j}(R_j|\lambda)$ be the outcome function that gives the probability of obtaining $R_j$ on measuring the system in $\lambda$ at $t_i$. Also let $\gamma_i(\lambda'|Q_i,\lambda)$ denote the probability that on measuring $Q$ at time $t_i$ the ontic state $\lambda$ is changed to $\lambda'$. Consequently, the joint probability $P(Q_i,R_j)$ is obtained as:
\begin{align*}
P(Q_i,R_j)=\int d\lambda'\int d\lambda \mu(\lambda)\xi_i(Q_i|\lambda)\gamma_i(\lambda'|Q_i,\lambda)\xi'_j(R_j|\lambda')
\end{align*}
Under the assumption of \textit{NIM}, $\gamma_i(\lambda'|Q_i,\lambda)=\delta(\lambda'-\lambda)$. So the joint probability becomes:

\begin{align*}
P(Q_i,R_j)=\int{d\lambda \mu(\lambda)\xi_i(Q_i|\lambda)\xi'_j(R_j|\lambda)}
\end{align*}

and the correlation function is given by:
\begin{align*}
C_{ij}&=\int{d\lambda \mu(\lambda)\sum_{Q_i=\pm1}Q_i\xi_i(Q_i|\lambda)\sum_{R_j=\pm1}R_j\xi'_j(Q_j|\lambda)}\\
&=\int{d\lambda\mu(\lambda)\expval{Q_i}_{\lambda}\expval{R_j}_{\lambda}}\\
\end{align*}
Consequently, the CHSH expression is given by:
\begin{align*}
\Delta_{CHSH}=\int d\lambda\mu(\lambda)I(\lambda)
\end{align*}
where $I(\lambda)= \expval{Q_0}_{\lambda}(\expval{R_0}_{\lambda}+\expval{R_1}_{\lambda})+\expval{Q_1}_{\lambda}(\expval{R_0}_{\lambda}-\expval{R_1}_{\lambda}) \leqslant2$.

\begin{align*}
\therefore\int d\lambda\mu(\lambda)I(\lambda) \leqslant 2
\end{align*}

 This completes the proof of the temporal CHSH inequality in an ontic model framework. That is, any theory with an underlying ontological framework in which measurements do not disturb the ontic state of the system is bound to satisfy the inequality $\Delta_{CHSH}\leqslant 2$.\\

We now shift our focus to the case when intervening projective measurements are executed on the system under consideration, in which case the correlators are altered:

\begin{align*}
C_{\hat{a},\hat{b}}= \beta_n{(\hat{a}\vdot\hat{r}_1)}{(\hat{r}_n\vdot\hat{b})}
\end{align*}
This, on substitution, leads to the following CHSH expression:
\begin{align}
\Delta_{CHSH} &=C_{00}+C_{01}+C_{10}-C_{11}\\
&=\beta_n(\phi_{01}(\phi_{n0}+\phi_{n1})+\phi_{11}(\phi_{n0}-\phi_{n1}))
\end{align}

with $\phi_{01}=\hat{a}_0\vdot\hat{r}_1$, $\phi_{11}=\hat{a}_1\vdot\hat{r}_1$, $\phi_{n0}=\hat{r}_n\vdot\hat{b}_0$, $\phi_{n1}=\hat{r}_n\vdot\hat{b}_1$, where we have taken the measurements made by Alice and Bob to be along $\{\hat{a}_0,\hat{a}_1\}$ and $\{\hat{b}_0,\hat{b}_1\}$ respectively. As can be easily seen from the above expression, 
\begin{align*}
\max_{\phi_{01},\phi_{11}}\abs{\Delta_{CHSH}}=2\abs{\beta_n}\max (\abs{\phi_{n0}},\abs{\phi_{n1}})
\end{align*}
and since, $\abs{\beta_n}\leqslant 1$, it follows that $\abs{\Delta_{CHSH}}\leqslant 2$. In other words, intervening projective measurements ensure that the CHSH inequality is satisfied and invasiveness of the measurement performed by Alice ``apparently" disappears. This property remains valid for arbitrarily large $n$, which rules out the achievability  of super-classical correlations once an intervening measurement is performed - further intermediate measurements cannot be implemented that could possibly retrieve the lost correlations.

\subsection*{IV} \label{IV}
Finally, we focus on the game (\ref{s6}) that we formulated in a spatio-temporal setting. An apposite modification was invoked by conveniently incorporating a temporal separation between the measurements performed by Alice and Bob. In the conventional spatial CHSH setting, Alice chooses to perform a measurement on the qubit in her possession along one of the two directions $\hat{a}_0$ and $\hat{a}_1$, while Bob measures along one of the two directions $\hat{b}_0$ and $\hat{b}_1$ on his qubit. Both measurements are performed (usually) at the same time. In the new version that we have considered, we have assumed Alice to be performing her measurement at an earlier time $t_A$ and Bob at a later time $t_B$. We show here that this temporal separation does not alter the correlation between the outcomes obtained by Alice and Bob. Denoting ${C}_{\hat{a},\hat{b}}(t_1,t_2)$ as the corresponding correlator and labelling the respective outcomes as $r$ and $s$, we observe that:
\begin{align*}
{C'}_{\hat{a},\hat{b}}(t_1,t_2)&=\sum_{r,s=\pm1}rs\Tr(\rho_{AB}\chi_{a}^r\otimes\mathds{1})\Tr(\rho_{a}^r\mathds{1}\otimes\chi_{b}^s)\\
\end{align*}
Here $\chi_{a}^r=\dfrac{1}{2}[\mathds{1}+r\va{\sigma}\vdot\hat{a}]$, $\chi_{b}^s=\dfrac{1}{2}[\mathds{1}+s\va{\sigma}\vdot\hat{b}]$, and $\rho_{a}^r$ is the state of the biqubit system to which the initial state $\rho_{AB}$ collapses when Alice obtains the outcome $r$ upon measuring along $\hat{a}$.
\begin{align*}
\therefore \rho_{a}^r=\frac{(\chi_{a}^r\otimes\mathds{1})\rho_{AB}(\chi_{a}^r\otimes\mathds{1})}{\Tr(\rho_{AB}\chi_{a}^r\otimes\mathds{1})}
\end{align*}
Substituting in the expression for the correlator, we obtain:
\begin{align*}
{C'}_{\hat{a},\hat{b}}(t_1,t_2)&=\sum_{r,s=\pm1}rs\Tr((\chi_a^r\otimes\mathds{1})\rho_{AB}(\chi_a^r\otimes\chi_{b}^s))\\
&=\sum_{r,s=\pm1}rs\Tr(\rho_{AB}(\chi_{a}^r\otimes\chi_{b}^s))\\
&=\expval{\va{\sigma}\vdot\hat{a}\otimes\va{\sigma}\vdot\hat{b}}_{\rho_{AB}}
\end{align*}

What happens if we insert some intermediate measurements on Alice's subsystem? We show, for simplicity, the case when only a single intermediate measurement is performed along $\hat{c}$. The corresponding outcomes are labelled as $t$ and the projectors as $\chi_c^t=\dfrac{1}{2}[\mathds{1}+t\va{\sigma}\vdot\hat{c}]$. A straightforward calculation then yields:

\begin{align*}
C''_{\hat{a},\hat{b}}&=\sum_{r,s,t} rs\Tr(\rho\chi_a^r\otimes\mathds{1})\Tr(\rho_a^r\chi_c^t\otimes\mathds{1})\Tr(\rho_c^t\mathds{1}\otimes\chi_b^s)\\
&=\sum_{r,t}r\Tr(\rho\chi_a^r\otimes\mathds{1})\Tr(\rho_a^r\chi_c^t\otimes\mathds{1})\Tr(\rho_c^t\mathds{1}\otimes\va{\sigma}\vdot\hat{b})\\
&=\sum_{r,t}r\Tr(\rho\chi_a^r\otimes\mathds{1})\Tr(\rho_a^r\mathds{1}\otimes\va{\sigma}\vdot\hat{b})\\
&=\Tr(\rho\va{\sigma}\vdot\hat{a}\otimes\va{\sigma}\vdot\hat{b})\\
&=\expval{\va{\sigma}\vdot\hat{a}_i\otimes\va{\sigma}\vdot\hat{b}_j}_{\rho_{AB}}
\end{align*}
where $\rho_c^t$ is the biqubit state formed when $t$ is obtained as an outcome for the intermediate measurement.\\

That is, in this case too, the correlators do not suffer any change. It is also easy to convince oneself that the form remains invariant even when an arbitrary number of intermediate measurements are performed on Alice's subsystem. However, the correlators are substantially modified when the intermediate measurements are performed on Bob's subsystem. Once again, we demonstrate the case for a single intermediate measurement.

\begin{align*}
C''_{\hat{a},\hat{b}}&=\sum_{r,s,t} rs\Tr(\rho\chi_a^r\otimes\mathds{1})\Tr(\rho_a^r\mathds{1}\otimes\chi_c^t)\Tr(\rho_c^t\mathds{1}\otimes\chi_b^s)\\
&=\sum_{r,t}r\Tr(\rho\chi_a^r\otimes\mathds{1})\Tr(\rho_a^r\mathds{1}\otimes\chi_c^t)\Tr(\rho_c^t\mathds{1}\otimes\va{\sigma}\vdot\hat{b})\\
&=\sum_{r,t}r\Tr(\rho\chi_a^r\otimes\mathds{1})\Tr(\rho_a^r\mathds{1}\otimes\chi_c^t(\va{\sigma}\vdot\hat{b})\chi_c^t)\\
&=\Tr(\rho(\va{\sigma}\vdot\hat{a})\otimes\sum_t\chi_c^t(\va{\sigma}\vdot\hat{b})\chi_c^t)\\
&=(\hat{b}\vdot\hat{c})\Tr(\rho\va{\sigma}\vdot\hat{a}\otimes\va{\sigma}\vdot\hat{c})\\
&=(\hat{b}\vdot\hat{c})\expval{\va{\sigma}\vdot\hat{a}\otimes\va{\sigma}\vdot\hat{c}}_{\rho_{AB}}\\
&=(\hat{b}\vdot\hat{c})C_{\rho_{AB}}{(\hat{a},\hat{c})}
\end{align*}

Then the CHSH expression looks like:
\begin{align}
\Delta_{CHSH}=(\hat{b}_0\vdot\hat{c}+\hat{b}_1\vdot\hat{c})C{(\hat{a}_0,\hat{c})}
+(\hat{b}_0\vdot\hat{c}-\hat{b}_1\vdot\hat{c})C{(\hat{a}_1,\hat{c})}
\end{align}
from which $\abs{\Delta_{CHSH}}\leqslant 2$ follows.\\

When multiple intervening measurements are performed, we have:
\begin{align}
C''_{\hat{a},\hat{b}}=\kappa_{n}C{(\hat{a},\hat{c}_1)}
\end{align}
where $\kappa_{n}=\beta_n(\hat{c}_n\vdot\hat{b})$ and Bell's inequality is still satisfied (this follows, again, from the consideration that $\abs{\kappa}\leqslant 1$). \\

As has already been emphasized earlier, it is necessary to perform the intermediate measurement(s) on the second qubit (in the possession of Bob), if one wishes to suppress any kind of super-classical correlation. This feature can be nicely construed in the following manner. Once Alice has performed a measurement on her qubit, the resulting bipartite state becomes separable (whether or not it was so to begin with), and any further measurement performed on Alice's qubit does not have any effect whatsoever on the state of Bob's qubit. Consequently the original correlations (local or nonlocal) between the outcomes of Alice (already obtained) and the counterfactual (yet to be obtained) outcomes of Bob's measurement remain unaffected by any intervention on Alice's qubit following the completion of her own measurement. On the contrary, if any intermediate measurement is performed on Bob's qubit, it does tamper with the state of his qubit following the collapse effected by Alice's measurement on her qubit  at an earlier time. As a result, the correlations between Alice's and Bob's outcomes are subjected to a modification, but interestingly, the ensuing correlations are found to comply with the CHSH inequality.

\end{document}